\documentclass[a4paper,fleqn,usenatbib]{mnras}
\usepackage{newtxtext,newtxmath}
\usepackage[T1]{fontenc}
\usepackage{ae,aecompl}

\usepackage{graphicx}	
\usepackage{amsmath}	
\usepackage{amssymb}	
\usepackage[
]{hyperref}
\usepackage[usenames,dvipsnames]{xcolor}
\usepackage{url}

\newcommand{\udef}{\stackrel{\mathrm{def}}{=}}

\defcitealias{marchant:19}{M19}

\DeclareRobustCommand{\Eqref}[1]{Eq.~\ref{#1}}
\DeclareRobustCommand{\Figref}[1]{Fig.~\ref{#1}}

\DeclareRobustCommand{\Secref}[1]{Sec.~\ref{#1}}

\title[Sensitivity of the lower-edge of the PISN BH mass gap to time dependent convection]{Sensitivity of the lower-edge of the pair instability
  black hole mass gap to the treatment of time dependent convection}

\author[Renzo et al.]{M.~Renzo$^{1,2}$\thanks{\href{mailto:mrenzo@flatironinstitute.org}{mrenzo@flatironinstitute.org}}, R.~J.~Farmer$^{2}$,
  S.~Justham$^{3,4,2}$, S.~E.~de~Mink$^{5,2}$,
\newauthor{Y.~G\"otberg$^{6}$, P.~Marchant$^{7,8}$}\\
{$^1$ Center for Computational Astrophysics, Flatiron Institute, New York, NY 10010, USA}\\  
{$^2$ Anton Pannekoek Institute for Astronomy and Grappa, University of
  Amsterdam, NL-1090 GE Amsterdam, The Netherlands}\\
{$^3$ School of Astronomy \& Space Science, University of the Chinese Academy of Sciences, Beijing 100012, China}\\
{$^4$ National Astronomical Observatories, Chinese Academy of Sciences, Beijing
  100012, China}\\
{$^5$ Center for Astrophysics | Harvard \& Smithsonian, 60 Garden Street, Cambridge, MA 02138, USA}\\
{$^6$ The observatories of the Carnegie institution for science, 813 Santa Barbara St., Pasadena, CA 91101, USA}\\
{$^7$ Department of Physics and Astronomy, Northwestern University, 2145 Sheridan Road, Evanston, IL
  60208, USA}\\
{$^8$ Institute of Astrophysics, KU Leuven,Celestijnlaan 200D, 3001 Leuven, Belgium}\\
}

\date{}

\pubyear{2019}

\begin{document}
\label{firsxtpage}
\pagerange{\pageref{firstpage}--\pageref{lastpage}}
\maketitle

\begin{abstract}
   Gravitational-wave detections are now probing the black hole (BH) mass
   distribution, including the predicted pair-instability mass gap. These data
   require robust quantitative predictions, which are challenging to
   obtain. The most massive BH progenitors experience episodic mass ejections on timescales shorter than
   the convective turn-over timescale. This invalidates the steady-state
   assumption on which the classic mixing-length theory relies.
   We compare the final BH masses computed with two different versions of the stellar
   evolutionary code \texttt{MESA}: (i) using the default implementation of~\cite{paxton:18} and
   (ii) solving an additional equation accounting for the timescale for convective
   deceleration. In the second grid, where stronger convection develops during
   the pulses and carries part of the energy, we find weaker pulses. This leads to lower amounts of mass
   being ejected and thus higher final BH masses of up to $\sim$\,$5\,M_\odot$. The
   differences are much smaller for the progenitors which determine
   the maximum mass of BHs below the gap. This prediction is robust at
   $M_{\rm BH, max}\simeq 48\,M_\odot$, at least within the idealized context of this study.
   This is an encouraging indication that current models are robust enough for
   comparison with the present-day gravitational-wave detections. However, the large
   differences between individual models emphasize the importance of improving the
   treatment of convection in stellar models, especially in the light of the
   data anticipated from the third generation of gravitational wave detectors.
\end{abstract}


\section{Introduction}
\label{sec:intro}

One of the most challenging aspects of simulating the interior evolution of
stars is the treatment of convection \citep[e.g.,][]{renzini:87, arnett:18,
  buldgen:19}. The development of convective motion in highly stratified media
is an inherently multidimensional problem, which involves turbulence. Spherically
symmetric stellar models typically rely on the Mixing Length Theory
(MLT,~\citealt{bohmvitense:58}), which provides an averaged description of
subsonic, steady-state convection. Albeit with many well-known caveats, MLT is
often a sufficient description for the energy transport and chemical mixing
provided by convection. This is because the evolutionary timescale of a star is
typically much longer than the convective turnover timescale: within a single
timestep it is reasonable to assume that the steady state described by
MLT can be achieved in the convective layers.

However, some stars can
experience dynamical phases of evolution which are  
too short for convection to achieve the steady state described by
MLT. One relevant example is the calculation of the spectrum of
  asteroseismological pulsations for stars with convective envelopes
  \citep[e.g.,][]{unno:67, gough:77}.  Cases where the short
  evolutionary timescale might influence the stellar structure include
  the helium flash \citep[e.g.,][]{nomoto:77} and more generally any
  explosive thermonuclear ignition that can drive convection
  \citep[e.g.,][]{nomoto:84, nomoto:87, takahashi:13}, very late
  evolutionary phases of massive star evolution
  \citep[e.g.,][]{couch:15, chatzopoulos:16}, dynamically unstable
  mass transfer from a convective donor star
  \citep[e.g.,][]{paczynski:72, lauterborn:72}, and stellar explosions\linebreak
  \citep[e.g.,][]{couch:13}.  In these situations, the
time-dependence of convection can become important, and stellar
evolution calculations typically lack a first principles model
for the convective acceleration. Sometimes, the convective
acceleration due to buoyancy is limited to a fraction of the local
gravitational acceleration to prevent unphysically large accelerations
\citep[e.g.,][]{arnett:69,wood:74}.

Here, we focus on a timely example of a situation in
which the time dependence of convection can be important: the evolution of very massive stars experiencing pulsational pair instability
(PPI,~\citealt{fowler:64};\citealt{barkat:67}). Because of the large mass
required to encounter this instability, it is expected to be a rare phenomenon in nature, but
the recent detection of black holes (BH) with masses
$30\,M_\odot \lesssim M_\mathrm{BH}\lesssim 50\,M_\odot$ \citep{GWTC1} has driven the interest in
understanding the evolution of the most massive (stellar) BH progenitors. To
fully harvest the information carried by gravitational waves and use it to
constrain stellar evolution, we need to have robust stellar models and
characterize their sensitivity to uncertain ingredients
\citep[e.g.,][]{farmer:16, renzo:17, davis:19, farmer:19}.

Stars that develop helium (He) core masses exceeding $M_\mathrm{He}\gtrsim30\,M_\odot$
are predicted to encounter the PPI and shed significant amounts of mass
in subsequent pulsation episodes \citep[e.g.,][]{rakavy:67, yoshida:16, woosley:17,
  takahashi:18, marchant:19, leung:19, woosley:19}. The amount of mass lost in
these pulses, together with the previous wind mass loss, determines the mass
distribution of BHs formed. Increasing further
to $M_\mathrm{He}\gtrsim 60\,M_\odot$, the instability becomes so violent that
the entire star is disrupted in a pair-instability supernova \citep[PISN,
][]{barkat:67, fraley:68}, without leaving any compact remnant. For
$M_\mathrm{He}\gtrsim 135\,M_\odot$, all the energy released by the
thermonuclear explosion is used to photodisintegrate the newly formed nuclei,
instead of accelerating the stellar gas, and BH formation resumes \citep[e.g., ][]{bond:84}. Thus,
PISNe are expected to carve a gap in the BH mass distribution.

Numerical simulations of the PPI evolution require following
hydrodynamical phases in between phases of hydrostatic equilibrium.
This can be done alternating the use of two different codes
\citep[e.g.,][]{chatzopoulos:12, chatzopoulos:13, yoshida:16,  takahashi:18}, which
however limits the number of pulsational events that can be followed.
To the best of our knowledge, two hydrodynamic Lagrangian stellar
evolution codes can now follow the evolution of
such massive stars. \cite{woosley:17, woosley:19} presented the
first grids of stellar models computed with the KEPLER code
\citep{weaver:78}, building upon pre-existing 
models computed with the same code \citep[][]{woosley:02, woosley:07}. Recently,
\cite{marchant:19, farmer:19, renzo:20:PPI_CSM} and \cite{leung:19} used two different
implementations of hydrodynamics in the open-source code \texttt{MESA} \citep{paxton:11,paxton:13,paxton:15,paxton:18,
  paxton:19} to simulate the evolution of PPI.

Several authors have noted that the amount of mass lost is sensitive to the treatment of convection,
both before and during the pulses \citep[e.g.,][]{woosley:17, marchant:19, leung:19}. Here, we compare two grids of massive bare He core models to highlight the differences resulting
from variations in the treatment of time-dependent convection. In one of our grids, convection is
treated similarly to \citealt{paxton:18} and \citet{leung:19}, while the other grid follows the
approach used in \citet{marchant:19} (hereafter, \citetalias{marchant:19}) and \cite{farmer:19}. In
\Secref{sec:grid_comparison}, we present the BH masses from both grids. In \Secref{sec:examples}, we
illustrate the differences in internal structure using two pairs of example stellar models.
\Secref{sec:discussion} compares the two treatments of time-dependent convection adopted
  here to other implementations existing in the literature and summarizes the main limitations of this study. We discuss the implications of our
results in \Secref{sec:conclusion}.

We do not aim at solving a problem that has remained
in stellar astrophysics for several decades, but hope to stimulate
improvements in stellar evolution models that also account for the
time-dependent behavior of convective motion.

\section{Methods}

We use the open-source stellar evolution code \texttt{MESA} to simulate the evolution of
bare He cores at metallicity $Z=0.001$ with masses in the range
$25\,M_\odot\leq M_\mathrm{He}\leq 70\,M_\odot$. All our input files are available at
\href{http://cococubed.asu.edu/mesa_market/inlists.html}{http://cococubed.asu.edu/mesa\_market/inlists.html},
and our models are available at \href{https://zenodo.org/record/3406320}{doi:10.5281/zenodo.3406320}.
We track the energy generation with the 22-isotope nuclear reaction network
\texttt{approx21\_plus\_co56.net}. Slightly before the star becomes dynamically unstable,
i.e., the pressure weighted volumetric averaged adiabatic index approaches 4/3 \citep[e.g.,][]{stothers:99}
  \begin{equation}
    \label{eq:gamma1_avg}
    \langle\Gamma_1\rangle \udef \frac{\int \Gamma_1P\,d^3r}{\int P\,d^3r} \equiv \frac{\int \Gamma_1 \frac{P}{\rho}\,dm}{\int
    \frac{P}{\rho}\,dm} \lesssim \frac{4}{3}\ \ ,
\end{equation}
we
employ the HLLC Riemann solver in \texttt{MESA}\footnote{Conversely, \cite{leung:19} used the
  \texttt{MESA} implementation of artificial viscosity (see also \citealt{paxton:15}).}
\citep[][]{toro:94}, without relying on artificial viscosity to capture shocks.
After a dynamical pulse, if/once the core has recovered hydrostatic equilibrium,
we create a new stellar model of reduced mass with the entropy and chemical
profile of the bound material. We do not include any wind mass loss, although
the treatment of winds is known to influence the core structure of massive stars
\citep[][]{renzo:17}. Preliminary tests including wind mass loss showed the same
trends discussed here. The impact of uncertainties related to winds and other
input physics on our PPI models are studied in \citet{farmer:19}. Tests
to ensure the robustness of our models against spatial and temporal discretization
are discussed in (\citetalias{marchant:19}, \citealt{farmer:19}, \citealt{renzo:20:PPI_CSM}).
We refer the interested readers to \citetalias{marchant:19} for a full
description of our setup. Here, we focus only on the treatment of
convection.

\begin{figure*}
  \centering
  \includegraphics[width=\textwidth]{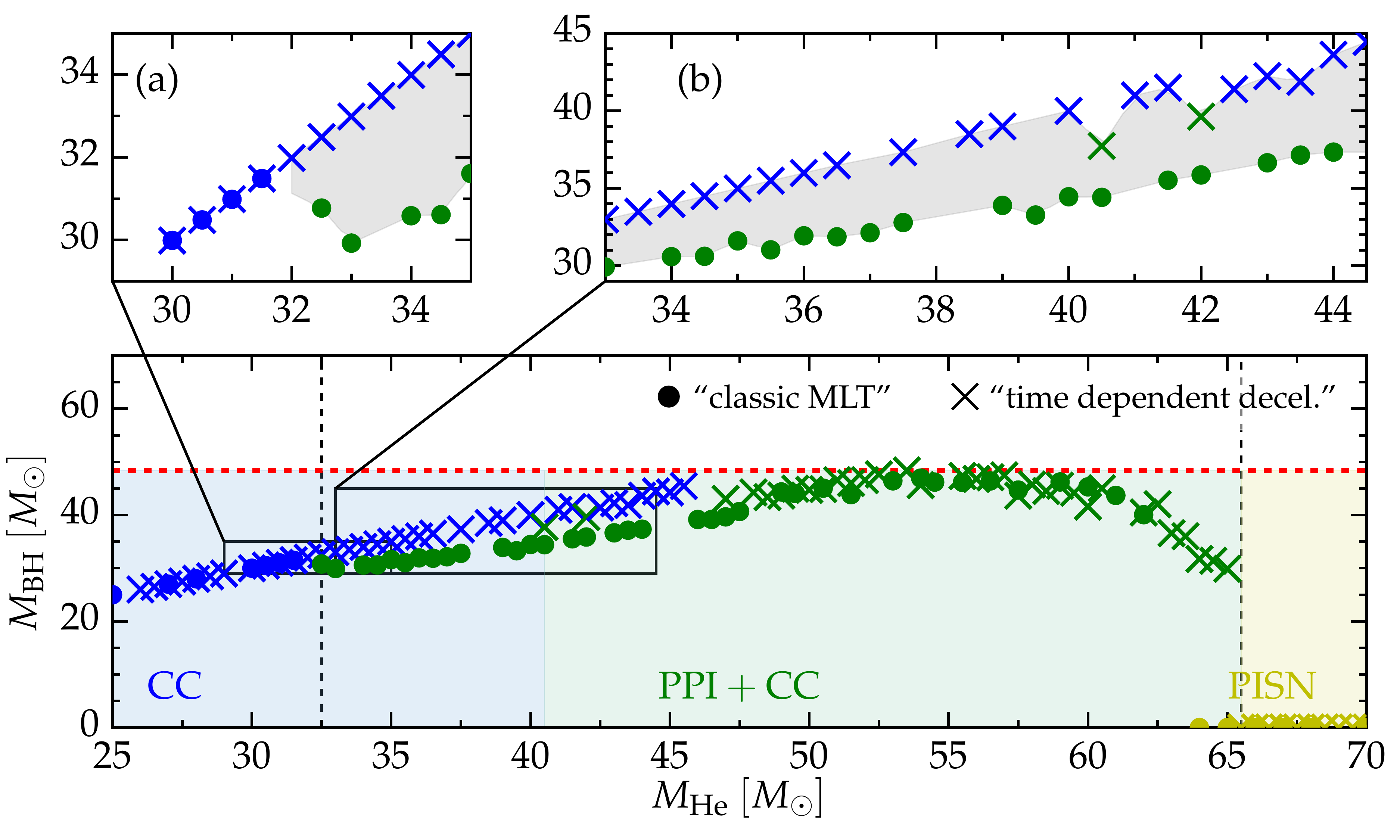}
  \caption{BH mass as a function of the He core mass for our two grids.
    The color shading indicates the approximate boundaries
    between evolution to core-collapse (CC, blue), pulsational pair-instability mass
    loss $\gtrsim3\,M_\odot$ (PPI+CC, green), and full
    disruption in a PISN (yellow). The gray area in the inset panels shows the
    systematic offset we find in the final BH masses from our two grids. The
    dashed red line indicates the maximum BH mass we find below the PISN BH mass gap
    (white area in the bottom panel), which is not sensitive to the variations between our two
    grids of models.}\label{fig:bh_dist}
\end{figure*}

We adopt the Ledoux criterion for convective
stability with a mixing length parameter
$\alpha_{\rm MLT}=2.0$ and an exponential under/overshooting with
(\texttt{f},\texttt{f\_0}) = (0.01,0.005) \citep[cf.~Eq.~2 in][]{paxton:11}.
To test the sensitivity of our results to the treatment of
time-dependent convection, we compute two grids of models using two different
\texttt{MESA} versions. Other differences between the two code versions might
contribute to the variations described here. Our first grid of models, which we
refer to as the ``classic MLT'' grid, is computed using \texttt{MESA}
release 10108. For this grid, the convective velocity $v_c$ is obtained from
MLT under the steady-state assumption, similarly to \cite{paxton:18} and
\cite{leung:19}, although the latter authors turn off convection during hydrodynamical phases of evolution. For
this grid we employ MLT++, which is an enhancement of the convective flux in superadiabatic radiation-pressure
dominated regions prone to developing density inversions \citep[][]{paxton:13,
  jiang:18} and a semiconvection efficiency of 0.01. We do not
  employ thermohaline mixing for numerical stability reasons. After the onset of the hydrodynamic phase of evolution we enforce
short timesteps, therefore we apply a limit to the convective acceleration based
on \cite{wood:74} to avoid unphysical infinite convective acceleration. This
approach still allows for infinite convective deceleration: if a stellar layer
becomes radiatively stable, the convective velocity is instantaneously set to zero. 

We compute our second grid, which we refer to as the ``time dependent deceleration''
grid, using  \texttt{MESA} version 11123. In this case, we obtain $v_c$ solving, together
with the stellar structure and composition equations, an equation 
designed to asymptotically give the MLT value of $v_c$ over long timescales, 
and to damp $v_c$ in radiative regions over a characteristic buoyancy
timescale. The equation we solve reads
(cf.~Eq.~A1 and A2 in \citetalias{marchant:19} and Eq.~11 in \citealt{arnett:69}):
\begin{equation}
  \label{eq:tdc_conv}
    \frac{\partial v_c}{\partial t}=  \begin{cases}
(v_{\rm MLT}^2 - v_c ^2)/\lambda  & \quad\rm{for\ convectively\ unstable\ regions} \quad , \\
-v_c^2/\lambda - Nv_c  & \quad\rm{for\ convectively\ stable\ regions} \qquad ,
 \end{cases}
\end{equation}
where $\lambda=\alpha_{\rm MLT} H_p$ is the mixing length, assumed to be proportional
through a free parameter $\alpha_\mathrm{MLT}$ to the
local pressure scale height $H_p$, $N$ is the Brunt-V\"ais\"al\"a frequency, and $v_\mathrm{MLT}$ is the MLT steady state
convective velocity. In this second ``time dependent
  deceleration'' grid, we do not employ MLT++. Semiconvection and
  thermohaline mixing are treated in the same way as in our ``classic
  MLT'' grid.

These two implementations of time dependent convection do not exhaust all the possible choices
 \citep[e.g.,][see also \Secref{sec:tdc_litterature}]{unno:67},
  but are sufficient to illustrate the qualitative and quantitative differences that can be
  expected in computing the evolution through PPI.
  
Our main parameter of interest is the resulting BH mass, which we estimate using
the mass coordinate where the gravitational binding energy reaches $10^{48}\,\mathrm{ergs}$. This allows for the
possibility of mass loss during the final core-collapse from either a weak
explosion \citep[][]{ott:18, kuroda:18}, energy loss to
neutrinos, or ejection of a fraction of
the envelope caused by the latter \citep[e.g.,][]{nadezhin:80,lovegrove:13}. This typically gives estimated BH masses within a few
$0.01\,M_\odot$ of the total baryonic mass slower than the escape velocity
at the onset of core collapse.

\section{Results}
\label{sec:results}


\subsection{Impact on the BH masses}
\label{sec:grid_comparison}

Figure~\ref{fig:bh_dist} shows the BH masses resulting from our
numerical experiment. Dots show models from our ``classic MLT'' grid, where increases in $v_c$ are
limited following \cite{wood:74} and the decreases in $v_c$ are unlimited, while
crosses mark the BH masses for models in our ``time dependent deceleration'' grid, which uses \Eqref{eq:tdc_conv}. The two inset panels
emphasize the main differences found, which could affect both the
the BH mass function and their detection rate in gravitational-wave events. 

The colors in \Figref{fig:bh_dist}
emphasize in blue the range of $M_\mathrm{He}$ that collapse without any
PPI-driven mass ejection (CC) and in green the PPI range, which we
define here requiring that PPI remove at least\footnote{Since we are concerned
  here with the features of the BH mass distribution, rather than all the potential
  observable signatures of a PPI, see also \cite{renzo:20:PPI_CSM}.} $3\,M_\odot$. The yellow region shows models
fully disrupted in a PISN. The boundary mass between PPI+CC behavior and full
disruption only shifts by $\sim$\,$3\,M_\odot$ between our two grids. This is smaller than variations
induced by other uncertainties (e.g., nuclear reaction rates and metallicity, \citealt{farmer:19}).

The inset (a) of \Figref{fig:bh_dist} magnifies the range at which PPI starts,
around $M_\mathrm{He}\simeq 32\,M_\odot$. This mass threshold for the occurrence of
thermonuclear explosions driven by the pair instability is in very good
agreement with \cite{woosley:17, woosley:19}. The models from our ``time dependent
convective deceleration'' grid (crosses) show, in this mass range, a one-to-one linear
correspondence between $M_\mathrm{He}$ and the BH mass. The occurrence of weak
pulses does not drive significant mass loss, blurring the boundary between CC
and PPI+CC evolution. Instead, the approach used in our ``classic MLT'' grid
produces stronger pulses at the low mass end, resulting in a turn-over in
$M_\mathrm{BH}\equiv M_\mathrm{BH}(M_\mathrm{He})$. Since lower mass He cores are
expected to be more common, if the pulses of the least massive stars
experiencing PPI can remove a significant amount of mass, then it might be
possible to detect an overabundance of BHs of mass corresponding roughly to the
minimum $M_\mathrm{He}$ for PPI.

The different amount of PPI mass loss for $M_\mathrm{He}\lesssim45\,M_\odot$ results in a systematic offset
in the final BH masses of $\sim$\,$5\,M_\odot$, shown in the inset (b) of
\Figref{fig:bh_dist}, and highlighted by the gray background in both inset panels. Models in the 
``time dependent convective deceleration'' grid generally produce more massive BHs, i.e.,
weaker pulses. This offset might affect the
mass-dependent binary BH merger rate by changing which stars make BHs of a given
mass. 
At $M_\mathrm{He}\simeq45\,M_\odot$, the ``time dependent
deceleration'' grid shows hints of a turn-over qualitatively similar
to the one at 32$\,M_\odot$ for our ``classic MLT'' grid, cf.~inset
(a). This feature might produce a concentration of BHs at the
corresponding BH mass
$M_\mathrm{BH}\simeq43\,M_\odot$. 

The PISN BH mass gap (the first part of which is shown by the white
region in
\Figref{fig:bh_dist}) starts above  $M_\mathrm{BH}\simeq48\,M_\odot$ for both our grids, also
in agreement with \cite{woosley:17,woosley:19}. The different treatment of
time-dependent convection in our two grids does not change the maximum BH mass
below the PISN BH mass gap significantly (red dashed line in \Figref{fig:bh_dist}),
corroborating the results of \cite{farmer:19}. For $M_\mathrm{He}\gtrsim45\,M_\odot$ the
scatter in BH masses increases, owing to the
combination of more energetic pulses and the lack of
wind mass loss in both our grids. The lack of winds, a situation possibly
  relevant for zero-metallicity population III stars, produces 
structures with sharp density drops: these influence the propagation of
shocks in the star and the amount of mass they remove. From a
computational perspective they result in numerically less stable models. Wind mass
loss (indirectly) and multi-dimensional effects are likely to smooth these
boundaries in nature. 

\subsection{Illustrative examples}
\label{sec:examples}

\begin{figure} 
  \centering
  \includegraphics[width=0.5\textwidth]{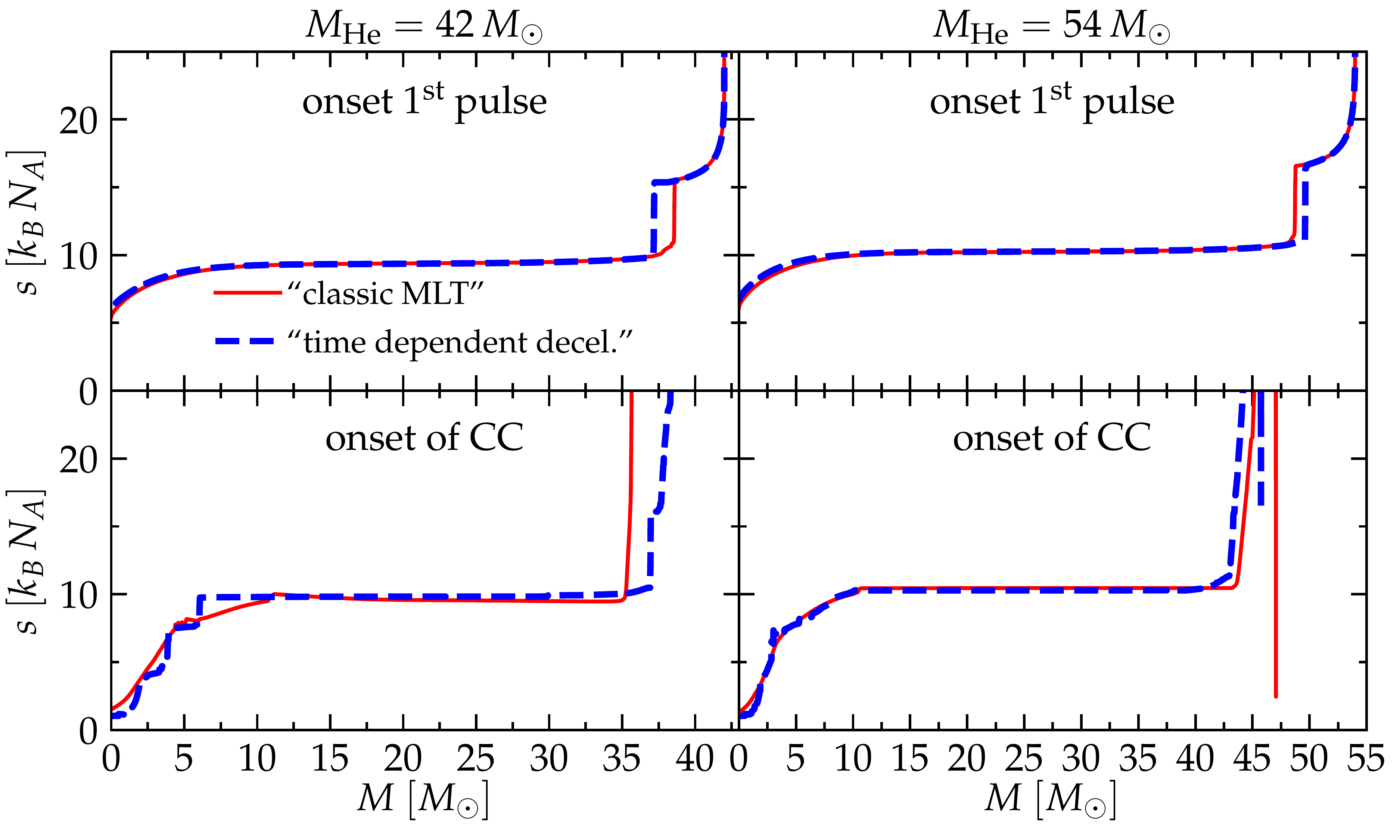}
  \caption{Specific entropy as a function of mass when the average adiabatic index
    $\langle \Gamma_1\rangle$ approaches 4/3 (\emph{top row}) and at the onset of the final
    core-collapse (\emph{bottom row}). Red solid lines are models from our ``classic MLT''
      grid, while thick blue dashed lines are models from our ``time dependent deceleration'' grid.
      We show a $42\,M_\odot$ model representative of the behavior of in the insets in
      \Figref{fig:bh_dist} (\emph{left column}) and a $54\,M_\odot$ model representative of 
      the progenitors of the most massive BHs below the PISN mass gap (\emph{right column}).}
  \label{fig:entropy}
\end{figure}

\begin{figure*}
  \centering
  \includegraphics[width=0.49\textwidth]{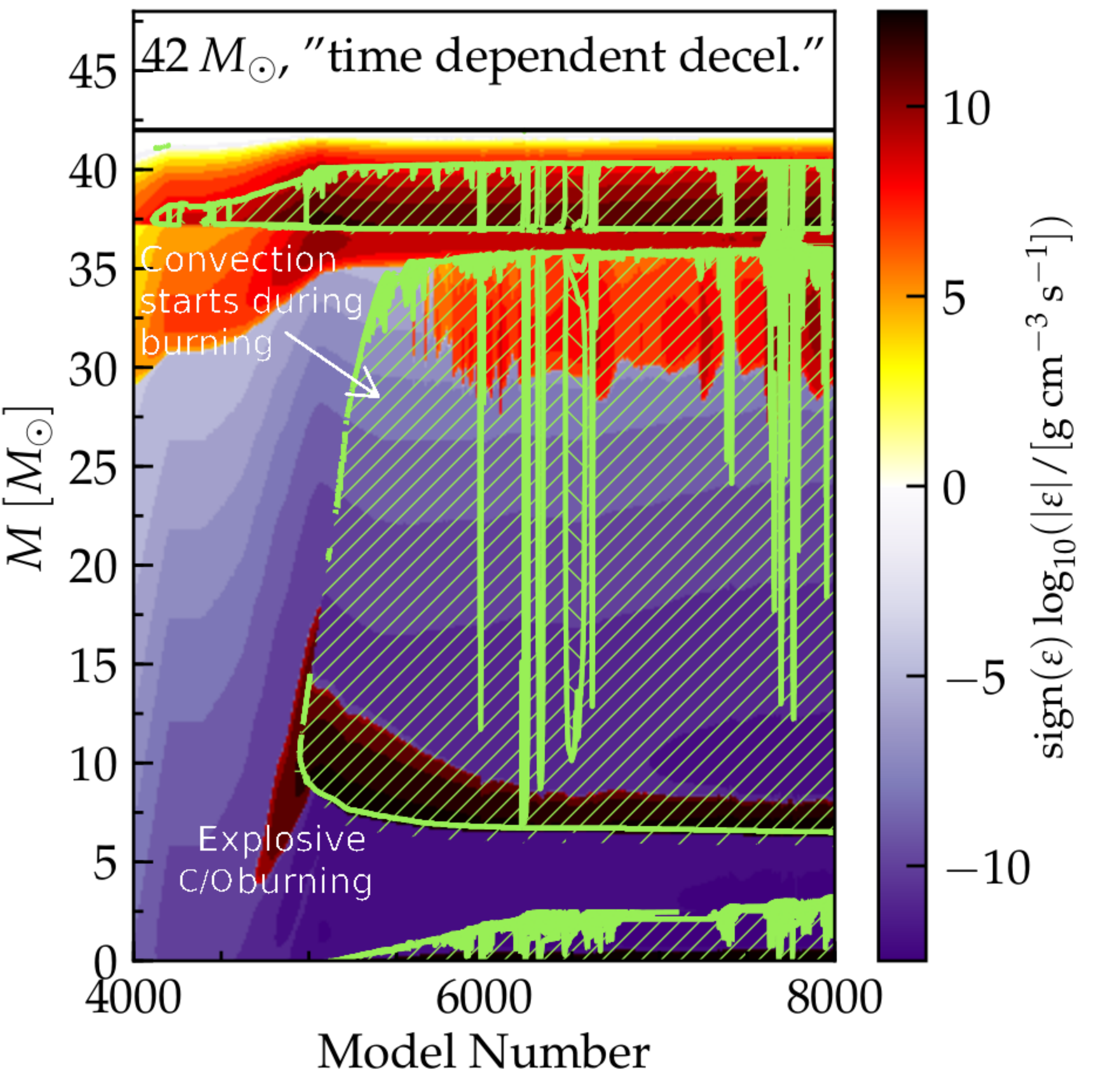}
  \includegraphics[width=0.49\textwidth]{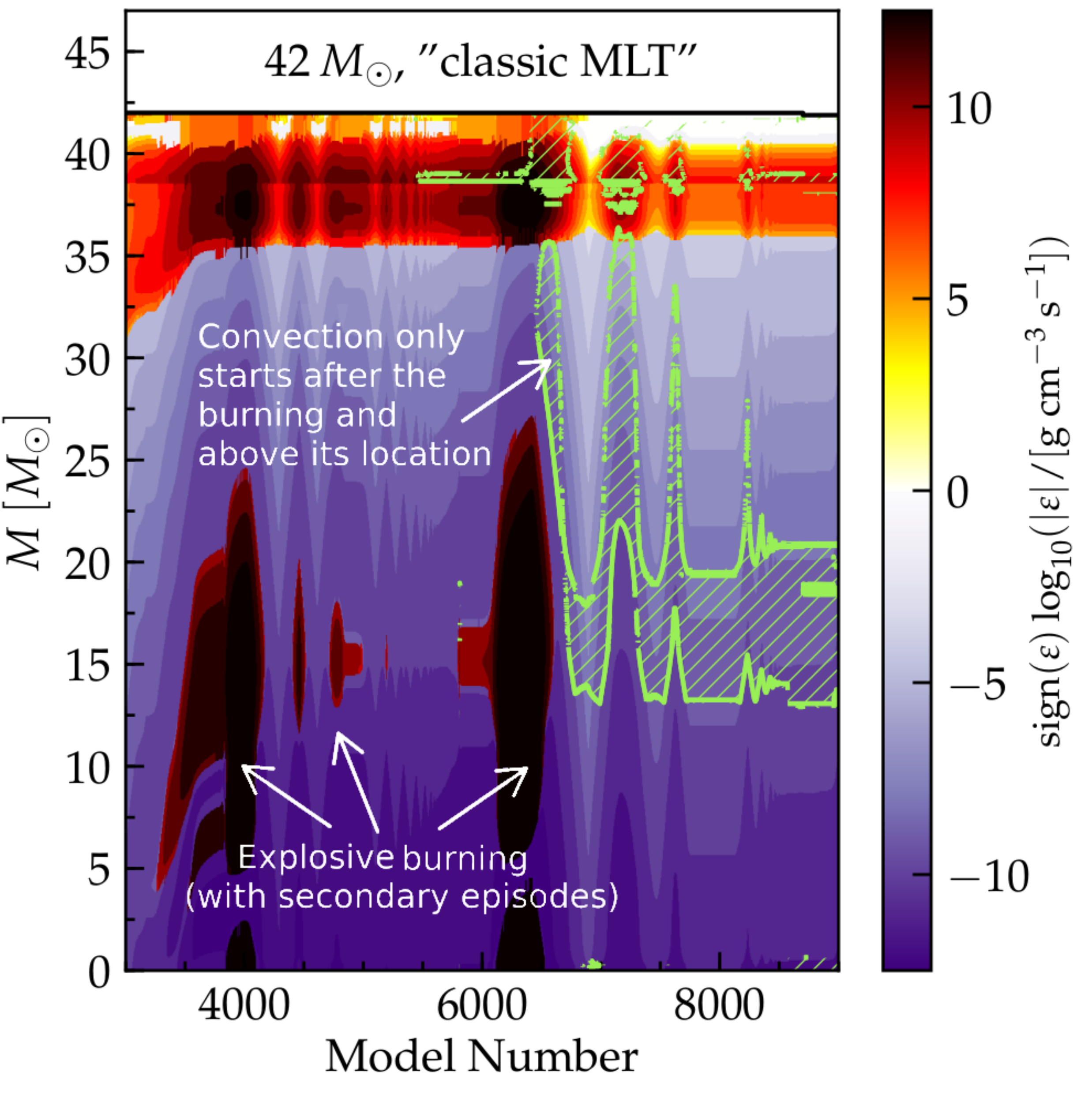}
  \includegraphics[width=0.49\textwidth]{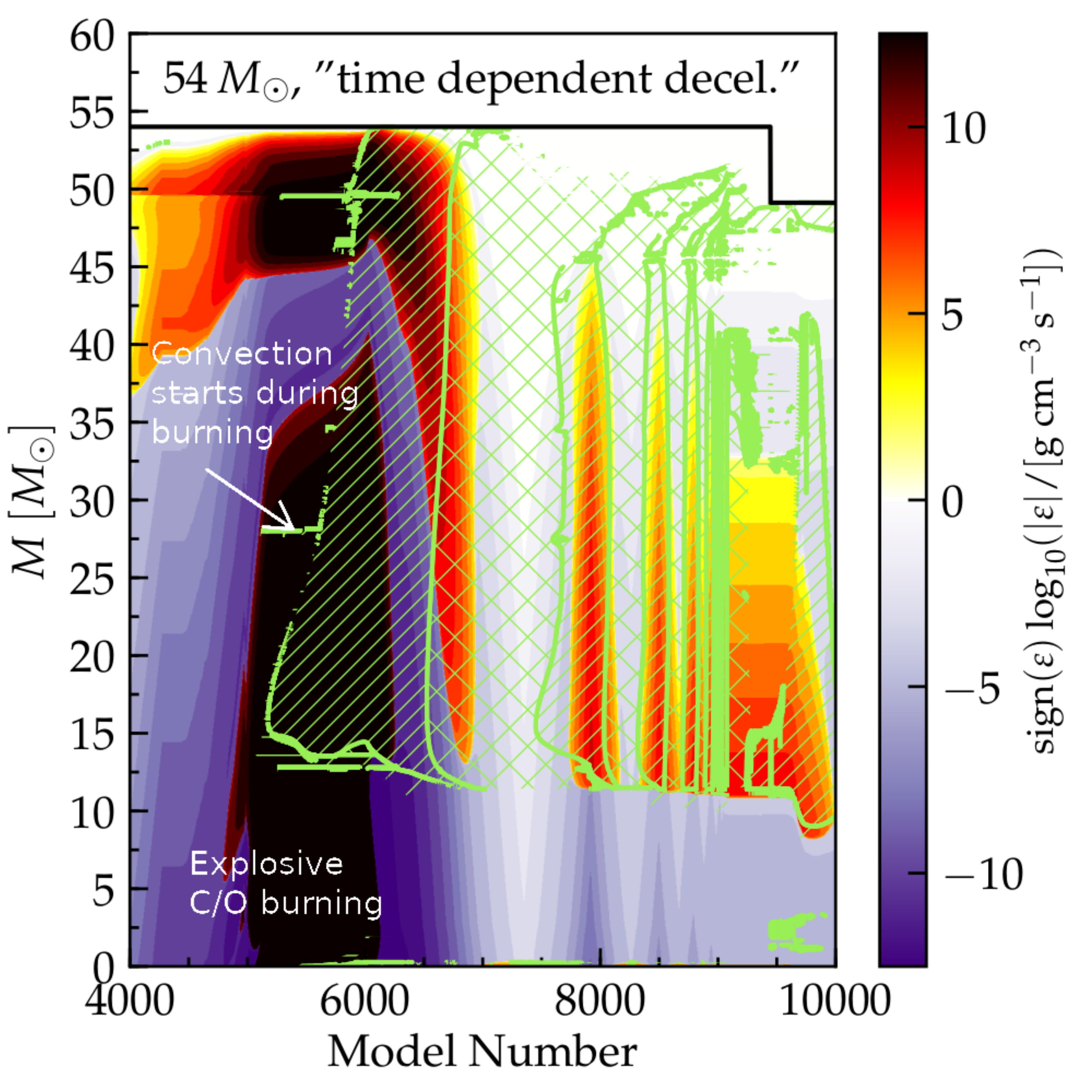}
  \includegraphics[width=0.49\textwidth]{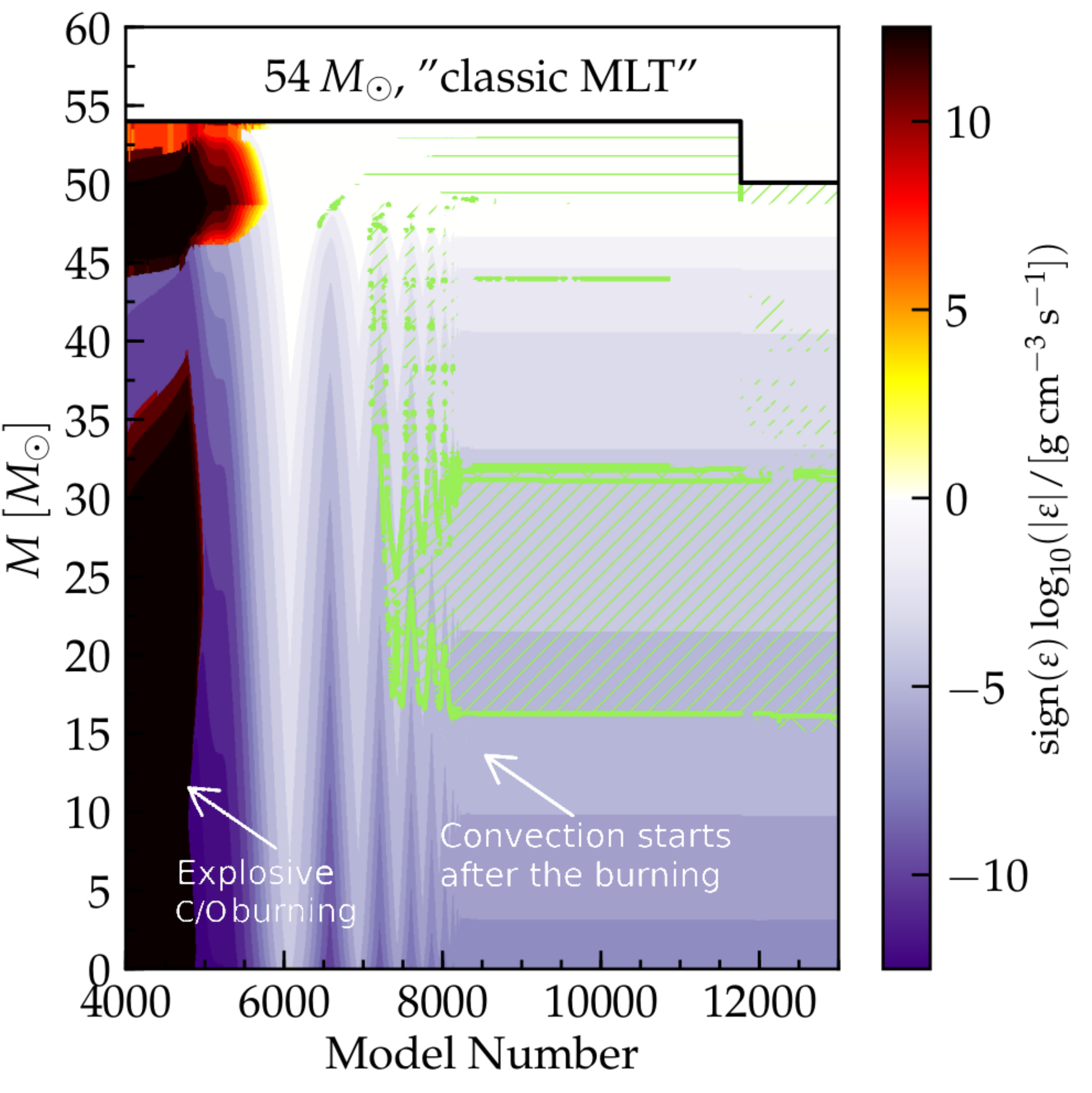}
  \caption{Kippenhahn diagrams during the development and propagation of the first pulse for the $42\,M_\odot$ (\emph{top row}) and $54\,M_\odot$
      (\emph{bottom row}) shown in \Figref{fig:entropy}. The left panels show the evolution for our ``time dependent deceleration''
    approach, while the right panels show the corresponding evolution for the
    ``classic MLT''. $\varepsilon=\varepsilon_\mathrm{nuc}-\varepsilon_\nu$ is the energy
    generation rate from nuclear burning minus the neutrino losses, there is net energy release for
    the red and black colors, and net energy loss for purple and blue colors. The
    green hatching indicates convective layers. The solid black line indicates
    the total mass of the models. A fraction of it
    becomes unbound earlier than it is removed from the computational domain.}
  \label{fig:kipp}
\end{figure*}

To illustrate the different internal evolution of the stars in our
grids we focus here on two pairs of models, of $42\,M_\odot$ and
$54\,M_\odot$ respectively. The first pair is representative of models
in the insets of \Figref{fig:bh_dist}, the second pair is
representative of the progenitors of the most massive BHs below the
PISN gap.

Figure~\ref{fig:entropy} shows the specific entropy as a function of
mass coordinate for these models in the conventional units of
Boltzmann's constant $k_B$ times Avogadro's number $N_A$. The specific
entropy characterizes the thermodynamic state of the gas, and it is
therefore useful when discussing thermal instabilities such as
convection.  Flat entropy profiles are a signature of efficient
convection.

For these models, the internal evolution is similar until the onset of
the first pulse and results in similar pre-pulse entropy profiles (top
panels). The behavior of convective shells during the pulses (i.e.,
when the star evolves on a dynamical timescale) shows a consistent
difference in the two grids. This can lead to divergent evolution and
different entropy profiles and final BH masses (at the lower mass
end), or not be sufficient to cause large differences in the final
entropy profile or BH mass (for the more massive PPI progenitors). The
insets of \Figref{fig:bh_dist}, and the $42\,M_\odot$ models in left
panels of \Figref{fig:entropy} show examples where the pulses drive
significant differences. Conversely, the $54\,M_\odot$ models in the
right column of \Figref{fig:entropy} are an example where the
differences in the convective acceleration/deceleration are not
sufficient to cause a different final BH mass and entropy profile.

Figure~\ref{fig:kipp} shows the Kippenhahn diagrams for the
$42\,M_\odot$ (top) and $54\,M_\odot$ (bottom) example models in our
``time dependent deceleration'' (left column) and ``classic MLT''
(right column) grids. These illustrate the differences in convective
patterns for our two grids during the pulses.

The age of the stars at the start of the pulses differ by a few
thousand years, a difference that we do not consider significant given
other numerical differences unrelated to the treatment of convection
in our grids. The time range shown is larger by a factor of
$\sim$\,$30$ in the models from the ``time dependent deceleration''
(right column), but corresponds to a smaller number of timesteps.
This is because we can run with the hydrodynamics on for much longer
and take longer timesteps thanks to the improved numerical stability
of the ``time dependent deceleration'' grid (see also Appendix ~\ref{sec:time}).

The oxygen thermonuclear explosion during the first pulse (roughly
between model number $\sim$\,$3\,000-6\,000$) proceeds differently in
our two grids.  In the ``time dependent deceleration'' models (left),
the oxygen ignition triggers convective mixing (green hatched areas)
during the main burning episode. Convection remains in the
intermediate layers of the star until and beyond the ejection of
mass. Conversely, in the ``classic MLT'' models (right panels of
\Figref{fig:kipp}) the thermonuclear explosion of oxygen is entirely
radiative, and convection turns on only after the main burning episode
is over, as the pulse wave propagates outward. As the core readjusts
dynamically to the energy released, secondary burning episodes are
clearly visible in the $42\,M_\odot$ model in the top right of
\Figref{fig:kipp}.

We think that the different development of convection occurs because
at its onset, computational zones of the models tend to oscillate
between radiative stability and convective instability. In the
``classic MLT'' grid, where infinite convective deceleration is
allowed, the convective velocity of such zones is reset to zero each
time they oscillate back to radiative stability. Conversely, in the
``time dependent deceleration'' grid, such zones retain a non-zero
convective velocity. The instantaneous value of the energy flux can
thus vary between these two approaches, and this creates a numerical
degeneracy between the convective deceleration, the onset of
convection, and the mixing processes happening at the boundary of the
convective regions (including the treatment of undershooting,
semiconvection, and thermohaline mixing).

At the lower mass end of the PPI regime (for $M_\mathrm{He}\lesssim
45\,M_\odot$), in the ``time dependent deceleration'' grid, the
presence of convection during the main burning episode leads to
efficient outward transport of energy and until it can eventually be
radiated away. Conversely, in the the ``classic MLT'' grid (left
column of \Figref{fig:kipp}), where convection does not develop as
promptly, the energy released in the thermonuclear explosion remains
trapped in the star. Ultimately, this energy contributes to the
kinetic energy of the gas and results in stronger mass ejections,
producing the kink in inset (a) of \Figref{fig:bh_dist} and the offset
in inset (b).

The development of large convective shell during the outward
propagation of a pulse in the ``time dependent deceleration'' grids
can lead to the injection of helium into the hotter and deeper
regions, and consequently to a large increase in nuclear energy
generation rate within the convective region. However the evolutionary
timescale is set by the dynamical propagation of the pulse, and it is
much shorter than the nuclear timescale. Therefore, while this burning
changes the chemical profile inside the star, it does not release an
amount of energy sufficient to modify significantly the dynamics of
the pulse propagation. Indeed, the amount of mass lost by both our
$54\,M_\odot$ models (bottom row of \Figref{fig:kipp}) in the first
pulse is similar, at about $4\,M_\odot$.

The lack of differences at the high mass end happens because the more
massive progenitors experience fewer but more energetic pulses
\citep[][]{woosley:17, marchant:19, renzo:20:PPI_CSM} and these are strong enough to
develop and sustain convection regardless of which algorithm is
used. The two algorithms we compare differ for how convection develops
(and damps), but once convection is going on they yield similar
results. In either case, the energy released by the thermonuclear
explosion during a pulse exceeds the amount that convection carries
away. The remaining differences in the resulting BH masses are smaller
than those introduced by other physical uncertainties (e.g., nuclear
reaction rates and overshooting, \citealt{farmer:19}).

Consequently, the maximum BH mass below the PISN BH mass gap, which is
produced by models at the more massive end (cf.~\Figref{fig:bh_dist}),
is robustly predicted at $\sim$$48\,M_\odot$ for models that do not
experience wind mass loss \citep[see also][]{farmer:19}, and does not
depend on the treatment of the convective deceleration, at least
within the framework of our comparison.

\section{Discussion}
\label{sec:discussion}

\subsection{Other time-dependent treatments of convection}
\label{sec:tdc_litterature}

The treatment of convection is a computationally challenging aspect of
stellar evolution \citep[e.g.,][]{renzini:87}. The specific aspect we
focus on here is its time-dependence, which becomes important when the
timescale of interest for the star is comparable to, or shorter than,
the convective turnover timescale.

Efforts to include the time dependence of convection in stellar
evolution calculations can be divided in two categories: (i) those
trying to capture time-dependent perturbations on a pre-existing
steady state described by MLT, and (ii) those concerned with the
growth (or damping) of the convective instability from the radiative
equilibrium (or from the MLT steady state).  Examples of (i) are the
calculations of the eigen-spectrum of stars with convective envelopes
(e.g., \citealt{unno:67}, \citealt{gough:77}, see also Sec.~2 in
\citealt{paxton:19}). The algorithm developed by \cite{unno:67} in
this context has also been applied to the problem of the growth of the
convective instability \citep[e.g.,][]{nomoto:77,
  takahashi:13}. Examples of (ii) are the algorithms of \cite{wood:74}
and \cite{arnett:69} on which our calculations are based.

The importance of convection in the context of PPI evolution was
investigated first by \cite{fraley:68}, and has been underlined in
many studies since then \citep[e.g.,][]{woosley:17, leung:19,
  marchant:19, farmer:19}. Modern calculations of PPI evolution either
turn off convection during the hydrodynamic phase of evolution for
numerical stability \citep[e.g.,][]{leung:19}, implement a limit on
the convective acceleration based on the local gravitational
acceleration, leaving the convective deceleration unlimited (as in our
``classic MLT'' grid), or use an ad-hoc equation to solve for the
convective velocity (see \Eqref{eq:tdc_conv} that we use in our ``time
dependent deceleration'' grid, see also \citealt{marchant:19,
  farmer:19, renzo:20:PPI_CSM}). The last two approaches fall into the category (ii) of
dealing with how the steady state described by MLT develops, and they
differ mainly for the inclusion or lack of a timescale for the damping
of convection. Our numerical experiments show that changing the
convective deceleration leads in a different onset of the convective
instability in our models.

Models based on the first approach (i) implicitly assume for the
convective velocity field the MLT value and compute small and
time-dependent perturbations to it \citep[see also][]{gough:77}.  This
is in principle not applicable to the case of PPI evolution, where the
convective instability grows as the star evolves and the
``perturbation'' in the velocity is the convective velocity itself. A
comparison between these two classes of treatments is beyond the scope
of the present study, although it would be interesting given the use
of the \citealt{unno:67} algorithm in other cases where the relevant
problem is the growth of convection.

\subsection{Further caveats}
We have carried out several experiments to ensure the numerical
convergence with increasing spatial and temporal resolution of our PPI
models, as presented in appendix B of \cite{marchant:19}, in
\cite{farmer:19}, and appendix A of \cite{renzo:20:PPI_CSM}. Nevertheless, we cannot exclude that the treatment
of convection is numerically degenerate with other minor differences
in the two \texttt{MESA} versions we employ here.

Of particular concern is the entrainment of the bottom edge of the
He-burning shell. In some of our models, this shell moves downwards in
mass coordinate, increasing the entropy of the outer layers of the CO
core, i.e., decreasing the total amount of mass at low entropy. We
find no clear trends in what determines whether the burning shell
penetrates downwards or not: this can happen in models that do not
experience PPI mass loss, but also in models that later on undergo
pulses. This can lead to differences in the pre-pulse entropy profiles
which are not driven by the different treatment of the time dependence
of convection, since they develop during evolutionary phases when the
star evolves on a much longer timescale than the convective turnover
timescale.

We suspect that this behavior depends on the sharp density and
composition profiles we obtain in the absence of winds, and
differences in the numerical setup in the two \texttt{MESA} versions
we employ. We have also found that turning off convective
undershooting or increasing the efficiency of semiconvective
mixing\footnote{Sparser grids computed with the ``time dependent
  deceleration'' setup but no undershooting or with semiconvection
  efficiency of 1.0 are also available at \href{https://zenodo.org/record/3406320}{doi:10.5281/zenodo.3406320}.} can
also influence this behavior.  Models computed with the ``time
dependent deceleration'' setup but no undershooting result in BH
masses in between the values shown in \Figref{fig:bh_dist}.

Nevertheless, regardless of the behavior of the He shell, the
development of convection during the dynamical phase of a pulse always
resembles what shown in \Figref{fig:kipp}. We expect that the
improvements in the treatment of convection are the main reason for
the differences, but we caution that we cannot exclude that other
minor differences contribute. We hope that cleaner numerical
experiments with improved models for the development/damping of
convection will become possible.

Finally, we emphasize that our calculations and the resulting BH mass
in \Figref{fig:bh_dist} do not account for possible consequences of
binary interactions, as they are obtained from the evolution of single
He cores. Further work to assess how binarity can modify the evolution
through PPI is needed to interpret gravitational-wave events assuming
the isolated binary evolution scenario.

\section{Summary \& Conclusion}
\label{sec:conclusion}

The ongoing search for gravitational waves is starting to provide
direct constraints on the BH mass function and probe the theoretically
predicted PISN BH mass gap \citep{fishbach:17,LIGOpop,
  stevenson:19}. This offers an unprecedented tool to understand the
physics of their massive star progenitors. This requires quantitative
predictions from stellar models robust enough for a sensible
confrontation with the data. The variations in the model predictions
resulting from algorithmic choices or simplifying assumptions should
be small compared to the input physics that we wish to test and the
observational uncertainties.

We have compared the predictions for the final BH masses at the lower
edge of the predicted mass gap computed with two different versions
and setups of the stellar evolutionary code \texttt{MESA}, which
differ primarily in the treatment of the time dependence of
convection. Our ``classic MLT'' grid adopts the defaults of
\cite{paxton:18}, while in our ``time dependent deceleration'' grid we
solve an additional equation incorporating the timescale for the
damping of convection.  Different groups have recently used setups
very similar to the two options we compare here
(\citetalias{marchant:19}, \citealt{leung:19}, \citealt{farmer:19}, \citealt{renzo:20:PPI_CSM}).

We find systematic differences when comparing individual models for
the same initial mass. The final BH masses computed with our
time-dependent treatment of convective deceleration are lower by up to
$\sim$\,$5\,M_\odot$ than those in our grid computed adopting the
classic mixing length theory. After inspection of the evolution of the
internal structure, this seems to be a consequence of more prompt and
stronger convection during the propagation of a pulse. Convection
carries out part of the energy, preventing it from becoming bulk
kinetic energy of ejecta. This results in weaker pulses and a lower
amount of mass ejected. The differences are largest for models near
the lower end of the mass range for pair pulsations to occur
($32\,M_\odot\lesssim M_\mathrm{He}\lesssim45\,M_\odot$), but are less
important for the higher mass range ($45\,M_\odot\lesssim
M_\mathrm{He}\lesssim64\,M_\odot$), because in these models convection
develops regardless of the algorithm employed, but is never sufficient
to remove a large fraction of the energy released in the thermonuclear
explosion. Because of this, we find that the predicted maximum BH mass
for BHs below the gap is robust at $\sim$\,$48\,M_\odot$.

For now, the robustness of the prediction for the location of the edge
of the gap is encouraging. Even the variations we find between the
grids for individual masses are smaller than the typical uncertainties
on the individual BH masses inferred from gravitational-wave
detections.

For the future, our results should be taken as a warning. The
variations we find between individual models are substantial. The
constraints from gravitational-wave events will become increasingly
precise with more detections. Moreover, we can anticipate an
increasing number of events detected with high signal-to-noise ratio,
and thus more accurately determined parameters for the individual
BHs. This will increase the the robustness needed from the stellar
model predictions.

The treatment of convection will likely remain a multifaceted
challenge, of which the time-dependence is only one aspect,
complimentary to other well known issues, but there are several ways
forward. Multi-dimensional hydrodynamic simulations applicable to the
stellar regime can be used to derive a more realistic expressions that
can be included in stellar evolutionary codes
\citep[e.g.,][]{meakin:07, couch:13, couch:14, arnett:18, arnett:18b,
  yoshida:19}. As a first step, a physically-motivated expression for
the convective acceleration in the right hand side of
\Eqref{eq:tdc_conv} could be derived from the flow observed in
multi-dimensional hydrodynamic simulations, instead of the ad-hoc
parametrizations presently used.

The increasing number of gravitational-wave events detected will
provide a major motivation for further improving the progenitor
models. The anticipated capabilities of third-generation detectors are
particularly promising. These should be able to detect massive binary
BHs across all redshifts where significant star formation occured in
the Universe. They would enable us to probe the evolution of the BH
mass distribution as a function of redshift and uncover possible
detailed features in the shape of the mass distribution, which bears
the imprints of the physical processes that govern the lives of their
massive star progenitors.\\

{ \textbf{Acknowledgements:}~We thank M.~Cantiello, D.~Hendriks,
  E.~Laplace, and L.~van~Son for helpful discussions. MR, SJ, and SdM
  acknowledge funding by the European Union's Horizon 2020 research
  and innovation programme from the European Research Council (ERC),
  Grant agreement No.\ 715063), and by the Netherlands Organisation
  for Scientific Research (NWO) as part of the Vidi research program
  BinWaves with project number 639.042.728. RF is supported by the
  Netherlands Organisation for Scientific Research (NWO) through a top
  module 2 grant with project number 614.001.501 (PI de Mink). This
  research was supported in part by the National Science Foundation
  under Grant No. NSF PHY-1748958, and stimulated by discussions at
  the ``The Mysteries and Inner Workings of Massive Stars'' at
  KITP. Simulations were carried out on the Dutch national
  e-infrastructure (Cartesius, project number 16343) with the support
  of the SURF Cooperative.}

\newpage
\bibliographystyle{mnras}

\newpage
\appendix

\section{Evolutionary timescales post-pulse}
\label{sec:time}
\begin{figure}
  \centering
  \includegraphics[width=0.5\textwidth]{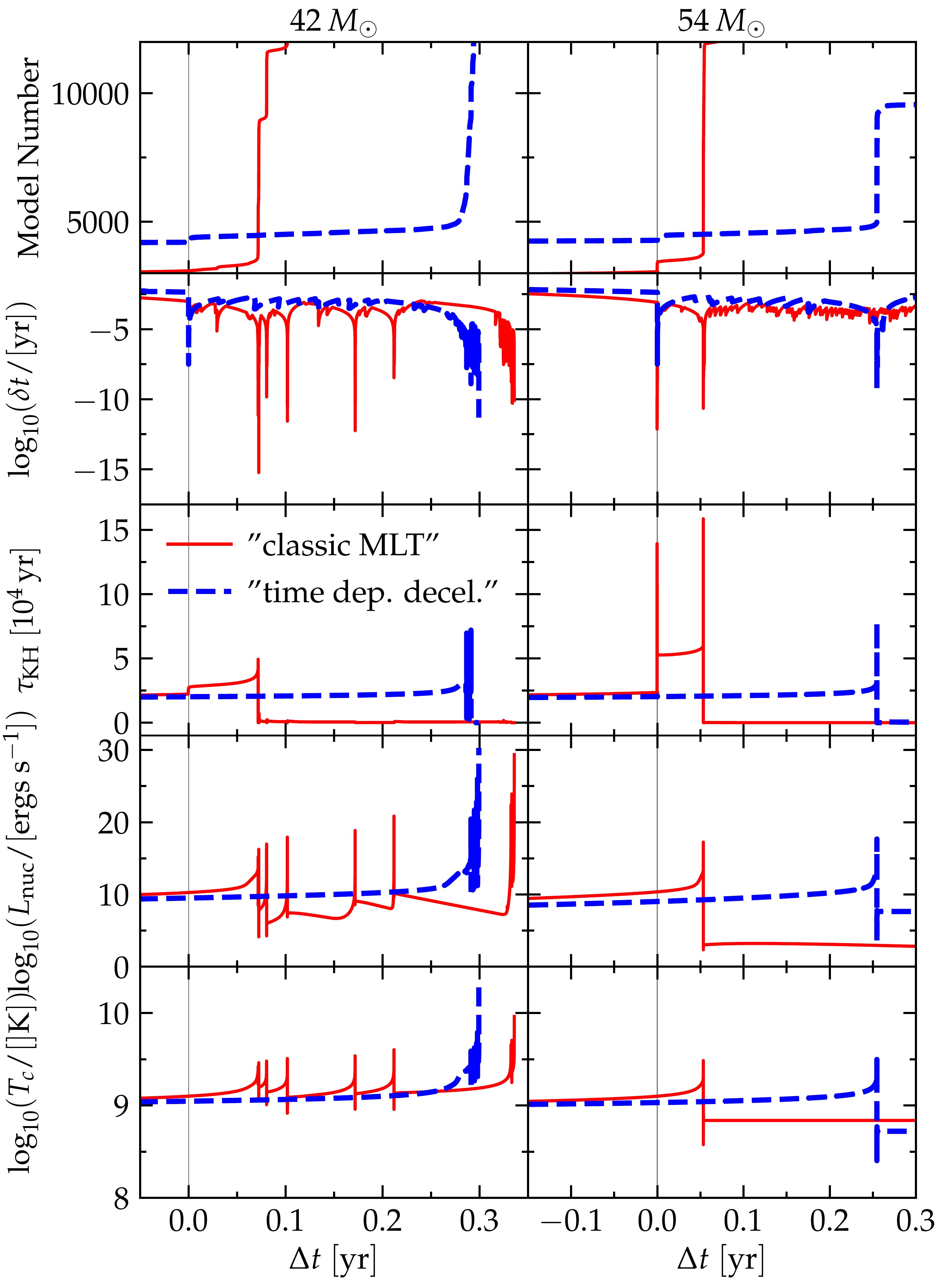}
  \caption{From top to bottom: evolution as a function of time
    $\Delta t$ since the onset of the dynamical
    instability (marked by the thin vertical line) of model number, timestep size
    $\delta t$, Kelvin-Helmholtz timescale, nuclear luminosity, and
    central temperature. Red solid curve show the ``classic MLT''
    models, while blue dashed curves show the ``time dependent
    deceleration'' models. The left (right) column shows the
    $42\,M_\odot$ ($54\,M_\odot$) pair of models.}\label{fig:timing}
\end{figure}

To clarify the onset of convection in our \texttt{MESA}
  models, \Figref{fig:kipp} shows the Kippenhahn diagrams using a
  non-physical quantity as x coordinate, namely the model
  number. Since our models are not computed with a fixed timestep,
  this makes it hard to read the amount of time elapsed. The total
  time elapsed corresponds to roughly a few tenths of a year, as shown in \Figref{fig:timing}.

\Figref{fig:timing} shows the evolution of the model number and
  timestep size $\delta t$ (top two panels), together with some global
  quantities (the Kelvin-Helmholtz timescale $\tau_\mathrm{KH}$ and
  nuclear luminosity $L_\mathrm{nuc}$
  integrated throughout the star, third and fourth panels from the
  top), and the central temperature $T_c$ (bottom panel) of the models
  shown in \Figref{fig:kipp}. These quantitites are shown as a function of the time elapsed since
  we turn on the HLLC solver (cf.~\Eqref{eq:gamma1_avg}), defined as $\Delta t=0$. As in
  \Figref{fig:entropy}, solid red curves show the ``classic MLT''
  models, and blue thicker dashed curves show the ``time dependent
  deceleration'' models. The left column shows the $42\,M_\odot$
  models, while the right column shows the $54\,M_\odot$ models
  (corresponding respectively to the top and bottom rows of
  \Figref{fig:kipp}). We chose the vertical and horizontal ranges to
  encompass the entire evolution shown in \Figref{fig:kipp} for all
  these models.

The Kelvin-Helmholtz timescale shown in the third panel is
  computed as a function of the total mass and radius of the models,
  however, these may include a significant amount of matter that is
  unbound and expanding rapidly to very large radii. At the
  onset of the pulses, the stars are evolving dynamically, on a much
  shorter timescale, and might temporarily be out of virial
  equilibrium because of the changing distribution of mass and
  consequently moment of inertia.

The spikes in $L_\mathrm{nuc}$ in the ``classic MLT''
  $42\,M_\odot$ models correspond to the secondary burning episodes
  highlighted in the top right panel of \Figref{fig:kipp}, and they
  correlate with a strong decrease in the timestep size and consequent
  increase of the model number around a given $\Delta t$. The use of
  \Eqref{eq:tdc_conv} in the ``time dependent deceleration models''
  allows us to use generally longer timesteps, turn on the HLLC
  earlier (more physical time elapses between $\Delta t=0$ and the
  first spike in $T_c$ corresponding a thermonuclear explosion), and keep
  it on for a longer physical time.

\end{document}